\documentclass[a4paper]{jpconf}

\usepackage{array}
\usepackage{braket}
\usepackage{slashed}
\usepackage{graphicx}
\usepackage{subfig}
\usepackage{amsmath, amssymb, amsfonts}
\usepackage{mathtools}
\usepackage{mathrsfs}
  \DeclareMathAlphabet{\mathpzc}{OT1}{pzc}{m}{it}
\usepackage{bm}

\usepackage{etoolbox}
\patchcmd{\thebibliography}{\advance\leftmargin\labelsep}
  {\labelsep=0.5cm \advance\leftmargin\labelsep}{}{}
  
\usepackage{textpos}
\usepackage{etoolbox}
\patchcmd{\thebibliography}{\advance\leftmargin\labelsep}
  {\labelsep=0.5cm \advance\leftmargin\labelsep}{}{}
  
\begin{document}

\title{Light neutrino masses from a\\ non-Hermitian Yukawa theory}
\author{J Alexandre\textsuperscript{1}, C M Bender\textsuperscript{2} and \underline{P Millington}\textsuperscript{3}}
\address{\textsuperscript{1}Department of Physics, King's College London,\\ London WC2R 2LS, United Kingdom \\
  \textsuperscript{2}Department of Physics, Washington University,\\ St. Louis, MO 63130, U.S.A.\\
  \textsuperscript{3}School of Physics and Astronomy, University of Nottingham,\\ Nottingham NG7 2RD, United Kingdom}
\ead{jean.alexandre@kcl.ac.uk, cmb@wustl.edu, p.millington@nottingham.ac.uk}

\begin{textblock}{4}(12,-9.2)
\begin{flushright}
\begin{footnotesize}
KCL-PH-TH/2017-10 \\
15 March 2017
\end{footnotesize}
\end{flushright}
\end{textblock}

\begin{abstract}
Working within the context of $\mathcal{PT}$-symmetric quantum mechanics, we begin by describing a non-Hermitian extension of QED that is both Lorentz invariant and consistent with unitarity. We show that the non-Hermitian Dirac mass matrix of this theory exhibits an exceptional point, corresponding to an effectively massless theory whose conserved current is either right- or left-chiral dominated. With this inspiration, we are able to construct a non-Hermitian model of light Dirac neutrino masses from Hermitian and anti-Hermitian Yukawa couplings that are both of order unity. We finish by highlighting potential phenomenological implications of this model.
\end{abstract}

\section{Introduction}

The observation of neutrino flavour oscillations has provided compelling empirical evidence that (at least two of) the neutrinos of the Standard Model (SM) of particle physics have nonzero masses~\cite{Olive:2016xmw}. Observations of the Cosmic Microwave Background (CMB) by ESA's Planck satellite, when combined with data from supernovae light curves and Baryon Acoustic Oscillations (BAO), put an upper limit on the sum of the SM neutrino masses of~\cite{Ade:2015xua}
\begin{equation}
\sum_j m_j\ <\ 0.23\ {\rm eV}
\end{equation}
at the $95\%$ confidence level. The masses of the SM neutrinos therefore appear to be ``unnaturally'' small compared with those of the other SM fermions.

Neutrino masses provide strong evidence for the presence of physics beyond the SM, and it is now widely accepted that the SM should be understood as a ``low-energy'' effective field theory (EFT), albeit an incredibly successful and highly predictive one. If we impose conservation of lepton number $({\rm L})$ on the SM EFT,  neutrinos must be of Dirac type and acquire their masses through a coupling to the SM Higgs field in the same way as the other SM fermions. In this case, we must stomach around 12 orders of magnitude difference between the top-quark and neutrino Yukawa couplings. Instead, if we maintain that conservation of lepton number is only accidental at low energies, the SM EFT contains a dimension-5 operator~\cite{Weinberg:1979sa} --- the Weinberg operator --- which violates lepton number by two units. After spontaneous symmetry breaking, this operator, schematically of the form $(L\phi)^2/\Lambda$, where $L$ is the charged-lepton doublet and $\phi$ is the Higgs doublet, gives rise to a Majorana mass term that is suppressed by the scale of new physics $\Lambda$. At tree-level, the Majorana mass term may be mediated by the exchange of heavy right-handed neutrinos (type I see-saw~\cite{Minkowski:1977sc, GellMann:1980vs, Yanagida:1979as,Mohapatra:1979ia,Schechter:1980gr}), by a scalar triplet (type II see-saw~\cite{Schechter:1980gr,Magg:1980ut,Cheng:1980qt,Lazarides:1980nt,Mohapatra:1980yp}) or a fermion $SU(2)_L$ triplet (type III see-saw~\cite{Foot:1988aq}). See-saw mechanisms are compatible with supersymmetric ultra-violet completions of the SM, and they can be realised naturally within grand unified theories (see, e.g.,~\cite{Deppisch:2015qwa} and references therein). Gauge-invariant fermion masses may also be generated radiatively at the three-loop level in $U(1)$ gauge theories by means of global anomalies mediated by scalar or pseudoscalar fields~\cite{Pilaftsis:2012hq}.  The presence of such pseudoscalar fields on manifolds with quantum torsion can lead to chirality-violating, right-handed Majorana mass terms generated gravitationally at the two-loop level by analogous anomalous operators~\cite{Mavromatos:2012cc}.

In addition to explaining the smallness of the light neutrino masses, the out-of-equilibrium decay of the heavy right-handed neutrinos in the early Universe can play a pivotal role in explaining the observed matter-anti-matter asymmetry in scenarios of leptogenesis~\cite{Fukugita:1986hr,Kuzmin:1985mm} (for reviews, see Refs.~\cite{Pilaftsis:1998pd,Buchmuller:2005eh,Davidson:2008bu,Blanchet:2012bk}). Therein, lepton number violating decays produce an initial excess in lepton number that is converted to a baryon number excess (${\rm B}$) via the ${\rm B}+{\rm L}$-violating sphaleron processes of the standard electroweak theory.

However, as theoretically appealing as the see-saw mechanism may be, the origin of the light neutrino masses is yet to be verified experimentally, and it is therefore prudent to consider alternatives. To this end, we recall that quantum field theories are often constructed with the following three properties in mind:
\begin{itemize}
\item [(i)] locality --- ensuring microcausality, i.e.~that the position-space canonical commutation relations (CCRs) vanish for space-like separations.

\item [(ii)] Lorentz invariance in vacuum.

\item [(iii)] Hermiticity --- guaranteeing real eigenvalues and unitary evolution.

\end{itemize}
A local, Lorentz-invariant theory with an Hermitian Hamiltonian is necessarily $\mathcal{CPT}$-invariant; this is the celebrated $\mathcal{CPT}$ theorem.

By sacrificing locality, it has been shown that neutrino oscillations can be accommodated  through $\mathcal{CPT}$-violating Dirac mass terms without enlarging the neutrino sector of the SM~\cite{Barenboim:2001ac}. Fermion masses can also arise dynamically in Lorentz-Invariance-Violating (LIV) $U(1)$ gauge theories, involving higher-order spatial derivatives of the gauge field~\cite{Alexandre:2011ux,Alexandre:2012dm}. Such models can emerge in the low-energy limit of quantum gravity theories~\cite{Mavromatos:2010ar}, and flavour-mixing interactions between the fermion and gauge fields may provide an alternative explanation for flavour oscillations~\cite{Alexandre:2013tya,Alexandre:2014zta}. The impact of non-Hermitian terms on neutrino oscillations in matter have also been considered~\cite{Ohlsson:2015xsa}. In the context of open quantum systems, it has been shown that dissipative effects can give rise to effective neutrino masses, leading to flavour oscillations for initially massless or mass-degenerate neutrinos~\cite{Benatti:2001fa}.  Since dissipative quantum systems can be described by non-Hermitian effective Hamiltonians, this suggests to consider possible modifications of the neutrino sector in which the constraint of Hermiticity is relaxed.

Our starting point is the following observation: \emph{All Hermitian matrices have real eigenvalues, but not all matrices with real eigenvalues are Hermitian.} Consider, for example, the complex (but non-Hermitian) $2\times 2$ matrix
\begin{equation}
\mathbf{A}\ =\ \begin{bmatrix} a+ib & c\\ c & a-ib\end{bmatrix}\,,
\end{equation}
where $a,b,c\in\mathbb{R}$. Its spectrum is
\begin{equation}
\lambda(\mathbf{A})\ =\ \big\{a\:\pm\:\sqrt{c^2\:-\:b^2}\big\}\;.
\end{equation}
There are three regimes:
\begin{itemize}

\item [(i)] When $|c|>|b|$, we are in the \emph{unbroken phase}, and the spectrum of $\mathbf{A}$ contains two distinct, real eigenvalues $\lambda(\mathbf{A}) = \big\{a\pm\sqrt{c^2-b^2}\big\}\subset\mathbb{R}$.

\item [(ii)] When $|c|<|b|$, we are in the \emph{broken phase}, and the spectrum of $\mathbf{A}$ comprises a complex-conjugate pair of eigenvalues $\lambda(\mathbf{A}) = \big\{a\pm i\,\sqrt{b^2-c^2}\big\}\subset\mathbb{C}$.

\item [(iii)] When $|c|=|b|\neq 0$, the eigenvalues merge, and we lose an eigenvector. We have one real eigenvalue $\lambda(\mathbf{A})=\big\{a\big\}\subset\mathbb{R}$, and the matrix $\mathbf{A}$ is defective, having the Jordan normal form
\begin{equation}
\mathbf{A}_J\ =\ \begin{bmatrix} a & 1  \\ 0 & a \end{bmatrix}\;.
\end{equation}
At $c=\pm\,b$, the eigenvalues exhibit a square-root singularity, and this is known as an \emph{exceptional point}.
\end{itemize}

In the context of Hamiltonian systems, Hermiticity is a sufficient but not necessary condition for obtaining real eigenvalues and unitary evolution. The latter can instead be guaranteed by the less restrictive and more physical constraint of $\mathcal{PT}$ symmetry, i.e.~symmetry of the Hamiltonian under the discrete space-time symmetries of parity ($\mathcal{P}$) and time-reversal ($\mathcal{T}$). It can then be shown that the evolution is unitary with respect to the positive semi-definite inner product $\braket{A|B}=A^{\mathcal{CPT}}\cdot B$ rather than the usual inner product $\braket{A|B}=A^{\dag}\cdot B$~\cite{Bender:2002vv}. Thus, for non-Hermitian $\mathcal{PT}$-symmetric quantum theories, Hermitian conjugation is superseded by $\mathcal{CPT}$ conjugation. A comprehensive introduction to $\mathcal{PT}$-symmetric quantum mechanics can be found in Ref.~\cite{Bender:2005tb}.

Consider again the matrix $\mathbf{A}$. Its form is equivalent to the Hamiltonian of a coupled system with gain and loss. The matrix $\mathbf{A}$ is not Hermitian, but it is $\mathcal{PT}$-symmetric: the parity transformation corresponds to interchanging the two systems, i.e.
\begin{equation}
\mathcal{P}:\mathbf{A}\ \longrightarrow\ \mathbf{A}^{\mathcal{P}}\ =\ \mathbf{P}\mathbf{A}\mathbf{P}^{\mathsf{T}}\;,\qquad \mathbf{P}\ =\ \begin{pmatrix} 0 & 1\\1 & 0\end{pmatrix}\;,
\end{equation}
and time-reversal is enacted by complex conjugation, interchanging gain and loss, i.e.
\begin{equation}
\mathcal{T}:\mathbf{A}\ \longrightarrow\ \mathbf{A}^{\mathcal{T}}\ =\ \mathbf{A}^{*}\;.
\end{equation}
When the coupling between the systems is strong ($|c|>|b|$), we are in a region of unbroken $\mathcal{PT}$ symmetry (the unbroken phase), and the eigenvectors of $\mathbf{A}$ are simultaneous eigenvectors of $\mathcal{PT}$. On the other hand, when the coupling is weak ($|c|<|b|$),  the eigenvectors of $\mathbf{A}$ are no longer eigenvectors of $\mathcal{PT}$, and the $\mathcal{PT}$ symmetry is broken (the broken phase). The $\mathcal{PT}$ phase transition occurs precisely at the exceptional point. Hermitian quantum field theories do not exhibit exceptional points. However, the radius of convergence of their perturbation series in powers of the coupling constant is precisely the distance in the complex coupling-constant plane to the nearest exceptional point~\cite{Bender:1969si}.

A non-Hermitian extension of quantum electrodynamics (QED) was first studied in Ref.~\cite{Bender:1999et} (see also Ref.~\cite{Milton:2003ax}), wherein the four-vector potential was taken to transform as an axial vector with an imaginary bare coupling to the $U(1)$ current. The $\mathcal{C}$ operator of this theory was constructed in Ref.~\cite{Bender:2005zz}, and it was shown that its $S$ matrix describes unitary evolution. The theory of a free Dirac fermion with an anti-Hermitian, parity-violating mass term was first introduced in Ref.~\cite{Bender:2005hf} and studied further in Ref.~\cite{Alexandre:2015oha}. The authors of Ref.~\cite{JonesSmith:2009wy} recognised that such an extension of the Dirac equation can maintain a nonzero mass matrix whilst simultaneously having massless dispersion relations, potentially allowing massless neutrinos to undergo flavour oscillations.  In what follows, we report the results of Ref.~\cite{Alexandre:2015kra}, expanding upon the suggestions made in Refs.~\cite{Alexandre:2015oha} and \cite{JonesSmith:2009wy} that such non-Hermitian mass matrices may have important implications for neutrino physics.

\section{Non-Hermitian extension of QED}

Following Ref.~\cite{Alexandre:2015kra}, we begin with a non-Hermitian extension of QED of the form\footnote{We do not suggest that this non-Hermitian extension of QED is realised in nature. Instead, it will act as a playground in which to study the behaviour of non-Hermitian gauge theories.}
\begin{equation}
\label{eq:QEDL}
\mathcal{L}\ =\ -\:\frac{1}{4}\,F_{\mu\nu}\,F^{\mu\nu}\:+\:\bar{\psi}(x)\big(i\gamma^{\mu}D_{\mu}\:-\:m\:-\:\mu\gamma^5\big)\psi(x)\:+\:\mathcal{L}_{\rm gf}\;,
\end{equation}
where $\psi(x)$ is a four-component Dirac fermion, $F_{\mu\nu}=\partial_{\mu}A_{\nu}\:-\:\partial_{\nu}A_{\mu}$ is the field-strength tensor, $\mathcal{L}_{\rm gf}$ is the gauge fixing term, and we include both vector and axial vector couplings to the $U(1)$ gauge field $A_{\mu}$:
\begin{equation}
D_{\mu}\ =\ \partial_{\mu}\:+\:i\big(g_V\:+\:g_A\gamma^5\big)A_{\mu}\;,
\end{equation}
where $g_V, g_A\in\mathbb{R}$. The Lagrangian contains both an Hermitian mass term $m\bar{\psi}\psi$ and an anti-Hermitian mass term $\mu\bar{\psi}\gamma^5\psi$. The latter changes sign on Hermitian conjugation by virtue of the anti-commuting nature of the gamma matrices, i.e.~$\big\{\gamma^{\mu},\gamma^5\big\}=0\ \forall\,\mu=0,1,2,3$. By the usual definitions of the discrete symmetry transformations on Fock space, that is the definitions appropriate for a theory with an Hermitian Hamiltonian, the anti-Hermitian mass term is $\mathcal{C}$ even, $\mathcal{P}$ odd, and $\mathcal{T}$ even. Thus, it appears $\mathcal{CP}$ and $\mathcal{CPT}$ odd. We recall that the usual \emph{Hermitian} term $\bar{\psi}i\gamma^5\psi$ is $\mathcal{C}$ even, $\mathcal{P}$ odd, and $\mathcal{T}$ odd.

In the massless limit $m=\mu=0$, the Lagrangian is invariant under the combined vector and axial vector gauge transformation
\begin{subequations}
\begin{gather}
A_{\mu}\ \longrightarrow\ A_{\mu}\:-\:\partial_{\mu}\phi\;,\\ \psi\ \longrightarrow\ \exp\big[i\big(g_V\:+\:g_A\gamma^5\big)\phi\big]\psi\;,\qquad \bar{\psi}\ \longrightarrow\ \bar{\psi}\exp\big[-i\big(g_V\:-\:g_A\gamma^5\big)\phi\big]\;.
\end{gather}
\end{subequations}

It can be shown~\cite{Alexandre:2015oha} that the Dirac equation is the one obtained by varying the Lagrangian with respect to $\bar{\psi}$ for fixed $\psi$. This fact is not obvious since one would obtain a different equation of motion by varying with respect to $\psi$ and taking the Dirac conjugate. Following the former procedure, we have
\begin{equation}
\big(i\gamma^{\mu}D_{\mu}\:-\:m\:-\:\mu\gamma^5\big)\psi(x)\ =\ 0\;.
\end{equation}
In order to obtain the dispersion relations, we can use the usual trick of acting again with the Dirac operator, yielding
\begin{equation}
\big(D^2\:-\:m^2\:+\:\mu^2\big)\psi(x)\ =\ 0\;.
\end{equation}
The energies are therefore given by
\begin{equation}
\omega^2\ =\ \mathbf{p}^2\:+\:M^2\;,\qquad M^2\ =\ m^2\:-\:\mu^2\;,
\end{equation}
which are real when $m^2\geq\mu^2$. We see that the dispersion relation is the same for particles and anti-particles, implying that there is no $\mathcal{CPT}$ violation.

The fact that the non-Hermitian extension of QED yields identical dispersion relations for particles and anti-particles is indicating that we should construct alternative definitions of the $\mathcal{C}$, $\mathcal{P}$ and $\mathcal{T}$ transformations that are appropriate to a non-Hermitian theory and under which this theory is $\mathcal{C}$ even, $\mathcal{P}$ odd and $\mathcal{T}$ odd~\cite{Bender:2005hf}. Under these transformations, the theory remains $\mathcal{CP}$ odd, but it is now $\mathcal{CPT}$ and $\mathcal{PT}$ even, thereby falling within the category of $\mathcal{PT}$-symmetric quantum field theories. The dynamical degrees of freedom for a non-Hermitian theory are not related by Hermitian (Dirac) conjugation, and it is for this reason that the correct equation of motion is obtained by varying with respect to $\bar{\psi}$ and not that obtained by taking the Dirac conjugate of the equation of motion found by varying with respect to $\psi$.

For comparison, in curved space-time, we can generate a similar Hermitian and locally $\mathcal{CPT}$-violating operator through coupling to the background curvature connection. This can be achieved, for example, in homogeneous anisotropic Bianchi Type II, VIII, and IX spacetimes for a locally rotating system~\cite{Debnath:2005wk} (see also Ref.~\cite{deCesare:2014dga}). One has
\begin{equation}
\mathcal{L}\ = \ \sqrt{-\,g}\;\bar{\psi}(x)\big(i\gamma^aD_a\:-\:m\big)\psi(x)\;,
\end{equation}
where $g=\mathrm{det}\,g_{\mu\nu}$ is the determinant of the curved space-time metric. The covariant derivative is given by
\begin{equation}
D_a\ \equiv\ \partial_a\:-\:\frac{i}{4}\,\omega_{bca}\,\sigma^{bc}\;,
\end{equation}
where
\begin{equation}
\omega_{bca}\ =\ e_{b\lambda}\big(\partial_ae^{\lambda}_c\:+\:\Gamma^{\lambda}_{\mu\nu}e^{\mu}_ce^{\nu}_a\big)\;,
\end{equation}
is the spin connection, $\sigma^{ab}=\tfrac{i}{2}\big[\gamma^a,\gamma^b\big]$ is the generator of tangent-space Lorentz transformations, $e_a^{\mu}$ is a vielbein and $\Gamma^{\lambda}_{\mu\nu}$ is the curved-space Christoffel symbol. We use lower-case Latin and Greek indices to identify the coordinates of the tangent and curved spaces, respectively. After some algebra, this Lagrangian can be rewritten in the form
\begin{equation}
\mathcal{L}\ =\ \sqrt{-\,g}\bar{\psi}(x)\big(i\gamma^a\partial_a\:-\:m\:-\:\gamma^5\gamma^aB_a\big)\psi(x)\;,
\end{equation}
where
\begin{equation}
B^d\ =\ \epsilon^{abcd}e_{b\lambda}\big(\partial_ae^{\lambda}_c\:+\:\Gamma^{\lambda}_{\nu\mu}e_c^{\nu}e_a^{\mu}\big)\;.
\end{equation}
If we consider $B^a$ to be a fixed external field, we have a Hermitian and $\mathcal{CPT}$-odd mass term $\gamma^5\gamma^aB_a$. In this case, the dispersion relations are
\begin{equation}
\big(\omega\:\mp\:B_0\big)^2\ =\ \big(\mathbf{p}\mp\mathbf{B}\big)^2\:+\:m^2\;,
\end{equation}
differing between particles and anti-particles.

The presence of the $\gamma^5$ in the anti-Hermitian mass term in Eq.~\eqref{eq:QEDL} indicates that we are treating the left and right chiralities differently. In fact, it is the left- and right-chiral sectors of this theory to which the analogy of a coupled system of gain and loss applies. Transforming to an explicit chiral basis, we have
\begin{equation}
\mathcal{L}\ =\ \begin{pmatrix} \psi_L^{\dag} & \psi_R^{\dag}\end{pmatrix}\begin{pmatrix} i\bar{\sigma}\cdot D_{-} & -\,m_+ \\ -\,m_- & i\sigma\cdot D_+\end{pmatrix}\begin{pmatrix} \psi_L \\ \psi_R\end{pmatrix}\;,
\end{equation}
where $\sigma^{\mu}=(\sigma^0,\sigma^i)$ and $\bar{\sigma}^{\mu}=(\sigma^0,-\,\sigma^i)$, $\sigma^i$ are the Pauli matrices, and
\begin{equation}
m_{\pm}\ =\ m\:\pm\:\mu\;,\qquad D^{\mu}_{\pm}\ =\ \partial^{\mu}\:+\:ig_{\pm}A^{\mu}\,\qquad g_{\pm}\ =\ g_V\:\pm\:g_A\;.
\end{equation}
The mass matrix has the form
\begin{equation}
\mathbf{m}\ =\ \begin{pmatrix} 0 & m_+ \\ m_- & 0 \end{pmatrix}\;.
\end{equation}
We see immediately that this matrix becomes defective in the limit $\mu\to\pm\,m$. For $\mu=+\,m$, we obtain a massless theory with chirality flips biased from left to right; for $\mu=-\,m$, we obtain a massless theory with chirality flips biased from right to left. The conserved current for this theory is
\begin{equation}
j^{\mu}\ =\ \bar{\psi}\gamma^{\mu}\bigg(1\:+\:\frac{\mu}{m}\,\gamma^5\bigg)\psi\ =\ \psi_L^{\dag}\bar{\sigma}^{\mu}\psi_L\bigg(1\:-\:\frac{\mu}{m}\bigg)\:+\:\psi_R^{\dag}\sigma^{\mu}\psi_R\bigg(1\:+\:\frac{\mu}{m}\bigg)\;.
\end{equation}
Thus, for $\mu=+\,m$, the left-chiral current decouples, and for $\mu=\,-\,m$, the right-chiral current decouples.

We can look at this another way. For $\mu=+\,m$, the Lagrangian has the form
\begin{equation}
\mathcal{L}\ =\ \psi_L^{\dag}i\bar{\sigma}\cdot D_-\psi_L\:+\:\psi_R^{\dag}i\sigma\cdot D_+\psi_R\:-\:2m\psi_L^{\dag}\psi_R\;,
\end{equation}
and we have the Weyl equations
\begin{equation}
i\sigma\cdot D_+\psi_R\ =\ 0\;,\qquad i\bar{\sigma}\cdot D_-\psi_L =\ 2m\psi_R\;.
\end{equation}
We can integrate out the left chirality, giving the on-shell Lagrangian
\begin{equation}
\mathcal{L}_{\rm on-shell}\ =\ \psi_R^{\dag}i\sigma\cdot D_+\psi_R\;,
\end{equation}
that of a massless, right-chiral Weyl fermion. Trivially, one can show that the full vector plus axial vector gauge invariance is recovered in this limit:
\begin{equation}
A_{\mu}\ \longrightarrow\ A_{\mu}\:-\:\partial_{\mu}\phi\;,\qquad \psi_R\ \longrightarrow\ \exp(ig_+\phi)\psi_R\;.
\end{equation}
Proceeding similarly for $\mu=-\,m$, we may integrate out the right chirality, giving the on-shell Lagrangian of a massless, left-chiral Weyl fermion, invariant under the gauge transformation
\begin{equation}
A_{\mu}\ \longrightarrow\ A_{\mu}\:-\:\partial_{\mu}\phi\;,\qquad \psi_L\ \longrightarrow\ \exp(ig_-\phi)\psi_L\;.
\end{equation}

It can also be shown (see Ref.~\cite{Alexandre:2015kra}) that this behaviour persists to all loop orders. For instance, in Feynman gauge, the one-loop photon polarisation tensor is given by
\begin{equation}
\Pi^{\mu\nu}(p)\ =\ -\,\frac{g_V^2\:+\:g_A^2}{2\pi^2}\big(p^{\mu}p^{\nu}-\eta^{\mu\nu}p^2\big)\big(B_{21}+B_1\big)\:+\:\frac{g_A^2}{\pi^2}\,\eta^{\mu\nu}M^2B_0\;,
\end{equation}
where $B_{0;1;21}\equiv B_{0;1;21}(p,M,M)$ are the Passarino-Veltman scalar form factors~\cite{Passarino:1978jh}. Thus,
\begin{equation}
p_{\mu}\Pi^{\mu\nu}(p)\ = \ \frac{g_A^2}{\pi^2}\,p^{\nu}M^2B_0\ \underset{\mu\ \to\ \pm\,m}{\longrightarrow}\ 0\;,
\end{equation}
and the Ward-Takahashi identity is recovered. Moreover, the right-left and left-right components of the one-loop vertex correction take the forms
\begin{subequations}
\begin{align}
\Lambda^\mu_{RL}\ &=\ \frac{g_+g_-}{4\pi^2}\,m_-\,\Big[(g_++g_-)\big(p^{\mu}\,
C_{11}+q^\mu\,C_{12}\big)\:+\:g_-(p^\mu+q^\mu)C_0\big]\;,\\[0.3em]
\Lambda^\mu_{LR}\ &=\ \frac{g_+g_-}{4\pi^2}\, m_+\,\Big[(g_++g_-) \big(p^\mu\,
C_{11}+q^\mu\,C_{12}\big)\:+\:g_+(p^\mu+q^\mu)C_0\Big]\;,
\end{align}\end{subequations}
again in terms of scalar form factors $C_{0;11;12}\equiv C_{0;11;12}(p,q,M,0,M)$~\cite{Passarino:1978jh}. We see that the vertex corrections vanish, respectively, in the limits $m=+\,\mu$ and $m=-\,\mu$, such that the structure of the mass matrix is preserved at the loop level.

\begin{figure}
\centering
\includegraphics[scale=0.27]{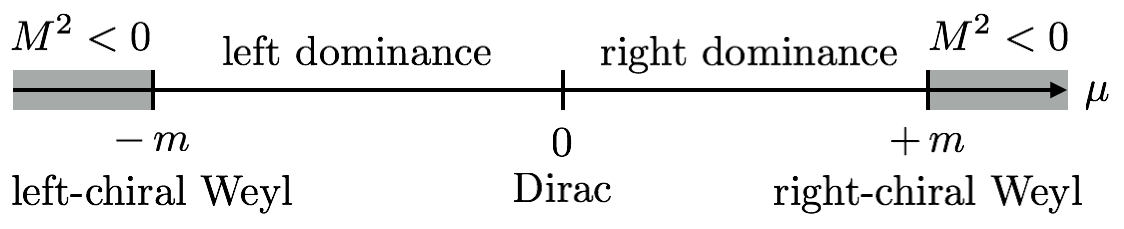}
\caption{\label{fig:phases} Schematic representation of the phases of the non-Hermitian models considered, indicating the continuum of theories that range from a massless, left-chiral Weyl theory (at $\mu=-\,m$) to a massless, right-chiral Weyl theory (at $\mu=+\,m$). The shaded regions indicate the tachyonic regimes, where $|\mu|>|m|$, giving $M^2<0$.}
\end{figure}

The behaviour of this theory is summarised graphically in Fig.~\ref{fig:phases}. By varying the anti-Hermitian mass term $\mu$, we can move smoothly from a massless, left-chiral Weyl theory (at $\mu=-\,m$) to a massless, right-chiral Weyl theory (at $\mu=+\,m$), recovering a massive Dirac fermion at $\mu=0$. For $-\,m\leq\mu<0$, the $U(1)$ current is dominated by the left chirality; for $0<\mu\leq+\,m$, the right chirality dominates. It is tempting to entertain the possibility that the light neutrinos of the SM might be described by a generalisation of the former case to the Higgs-Yukawa theory, and we will do so in the next section.

\section{Non-Hermitian Higgs-Yukawa theory}

Inspired by the behaviour of the non-Hermitian $U(1)$ theory described in the previous section, we now turn our attention to a one-generation, chiral Higgs-Yukawa theory with non-Hermitian Yukawa couplings:
\begin{equation}
\label{eq:YukLag}
\mathcal{L}\ = \ \bar{L}_Li\slashed{D}L_L\:+\:\bar{\nu}_Ri\slashed{\partial}\nu_R\:-\:h_-\bar{L}_L\tilde{\phi}\nu_R\:-\:h_+\bar{\nu}_R\tilde{\phi}^{\dag}L_L\;.
\end{equation}
Here, $L_L=(\nu_L\ e_L)^{\mathsf{T}}$ is the charged-lepton doublet, $\nu_R$ is a right-handed singlet, $\tilde{\phi}=i\sigma_2\phi^*$ is the isospin conjugate of the Higgs doublet and $D_{\mu}$ is the covariant derivative of the electroweak gauge groups of the SM.

We can emulate the behaviour of the non-Hermitian $U(1)$ theory of the preceding section by taking the non-Hermitian Yukawa couplings $h_+$ and $h_-$ to have the structure
\begin{equation}
h_{\pm}\ \equiv\ h\:\pm\:\eta\ \in\ \mathbb{R}\;.
\end{equation}
After spontaneous symmetry breaking and working in unitary gauge, the Higgs doublet and its isospin conjugate take the forms
\begin{equation}
\phi\ = \ \frac{1}{\sqrt{2}}\begin{pmatrix} 0\\ v+H\end{pmatrix}\;,\qquad \tilde{\phi}\ =\ \frac{1}{\sqrt{2}}\begin{pmatrix} v+H \\ 0\end{pmatrix}\;,
\end{equation}
where $v\sim 246\ {\rm GeV}$ is the vacuum expectation value of the Higgs field. Neglecting the $Z$- and Higgs-boson ($H$) couplings, the neutrino sector of the Lagrangian in Eq.~\eqref{eq:YukLag} becomes
\begin{equation}
\mathcal{L}\ \supset \ \bar{\nu}_L i\slashed{\partial}\nu_L\:+\:\bar{\nu}_Ri\slashed{\partial}\nu_R\:-\:m_-\bar{\nu}_L\nu_R\:-\:m_+\bar{\nu}_R\nu_L\;,
\end{equation}
where
\begin{equation}
m_{\pm}\ \equiv\ m\:\pm\:\mu\ =\ \frac{v}{\sqrt{2}}\,(h\pm\eta)\;.
\end{equation}
The non-Hermitian Yukawa couplings have given rise to Hermitian ($m=v h/\sqrt{2}$) and anti-Hermitian ($\mu=v\eta/\sqrt{2}$) masses that are analogous to those of the model in Eq.~\eqref{eq:QEDL}. For aesthetic reasons, we have changed $\mu\to -\,\mu$ relative to the preceding section. The resulting neutrino mass matrix has eigenvalues
\begin{equation}
\pm\,M\ =\ \pm\,\frac{v}{\sqrt{2}}\sqrt{h^2-\eta^2}\;,
\end{equation}
and we see that the neutrino becomes massless in the limit $h\to\pm\,\eta$. In fact, for $h\to\eta$, we obtain the original SM (with one generation): a theory of massless, left-handed neutrinos. On the other hand, for $\eta\lesssim h$ ($\eta>0$),\footnote{We correct an erroneous inequality appearing below Eq.~(4.11) in Ref.~\cite{Alexandre:2015kra}, which should instead read $0<\eta<h$.} we can obtain an arbitrarily small neutrino mass with the neutrino current still dominated by the left-chiral component. Most interestingly, these small masses can be generated for Hermitian and non-Hermitian Yukawa couplings that are both of order unity. We emphasise that the masses of the left- and right-handed neutrinos are degenerate at tree level in this construction.

For a single generation, we are restricted to having real-valued Yukawa couplings. To see this, let us suppose that
\begin{equation}
-\:\mathcal{L}\ \supset\ h_-\bar{L}_L\tilde{\phi}\nu_R\:+\:h_+^*\bar{\nu}_R\tilde{\phi}^{\dag}L_L\;,\qquad h_{\pm}\ \in\ \mathbb{C}\;.
\end{equation}
The mass eigenvalues are the roots of
\begin{equation}
M^2\ =\ \frac{v^2}{2}\Big(|h|^2-|\eta|^2-2i\mathrm{Im}\,h^*\eta\Big)\;,
\end{equation}
which can be real and nonzero iff $h$ and $\eta$ are real, distinct and nonzero.\footnote{We correct an erroneous factor of $1/2$ in Eq.~(4.15) of Ref.~\cite{Alexandre:2015kra}.} This is no longer the case if we extend to multiple flavours.

In the presence of more than one generation, the Yukawa couplings in Eq.~\eqref{eq:YukLag} are promoted to matrices in flavour space, and the relevant part of the Lagrangian becomes
\begin{equation}
\mathcal{L}\ \supset\ \bar{L}_{L,k}i\slashed{D}L_{L,k}\:+\:\bar{\nu}_{R,\alpha}i\slashed{\partial}\nu_{R,\alpha}\:-\:[h_-]_{k\alpha}\bar{L}_{L,k}\tilde{\phi}\nu_{R,\alpha}\:-\:[h_+^*]_{k\alpha}\bar{\nu}_{R,\alpha}\tilde{\phi}^{\dag}L_{L,k}\;.
\end{equation}
The left-handed and right-handed flavours are indexed by the lower-case Roman character $k$ and lower-case Greek character $\alpha$, respectively. If we take $N$ left- and $N$ right-handed flavours, the left-handed doublet $L_{L,k}$ and the right-handed singlet $\nu_{R,\alpha}$ transform in the fundamental representations of two flavour groups $U_L(N)$ and $U_R(N)$. 

For the case of two flavours ($N=2$) and after spontaneous symmetry breaking, the mass eigenvalues are given by the roots of
\begin{equation}
M_{1(2)}^2\ =\ \frac{v^2}{4}\Big[{\rm tr}\,\bm{h}_+^{\dag}\bm{h}_-\:-(+)\:\Big(2\,{\rm tr}\big(\bm{h}^{\dag}\bm{h}_-\big)^2\:-\:\big({\rm tr}\,\bm{h}_+^{\dag}\bm{h}_-\big)^2\Big)^{1/2}\Big]\;.
\end{equation}
It is clear that if $\bm{h}=\pm\,\bm{\eta}$, we obtain a massless spectrum. Instead, if $\mathrm{det}\,\bm{h}^{\dag}_+\bm{h}_-=0$, we obtain one massless ($M_1^2=0$) and one massive state:
\begin{equation}
M_2^2\ =\ \frac{v^2}{2}\,\mathrm{tr}\,{\bm h}^{\dag}_+{\bm h}_-\ =\ \frac{v^2}{2}\Big[{\rm tr}\,\bm{h}^{\dag}\bm{h}\:-\:{\rm tr}\,\bm{\eta}^{\dag}\bm{\eta}\:-\:2i\,{\rm Im}\,{\rm tr}\,\bm{h}^{\dag}\bm{\eta}\Big]\;.
\end{equation}
The mass is real and non-tachyonic so long as $\mathrm{Im}\,\mathrm{tr}\,\bm{h}^{\dag}\bm{\eta}=0$ and $\mathrm{tr}\,\bm{h}^{\dag}\bm{h}>\mathrm{tr}\,\bm{\eta}^{\dag}\bm{\eta}$. In this two-flavour case and by tuning the Yukawa couplings so that $\mathrm{tr}\,\bm{h}^{\dag}\bm{h}\gtrsim\mathrm{tr}\,\bm{\eta}^{\dag}\bm{\eta}$, the mass-splitting $\Delta M^2=M_2^2-M_1^2$ can be made arbitrarily small, whilst at the same time maintaining Hermitian and non-Hermitian Yukawa couplings of order unity.

We have so far considered only the Dirac masses. However, we are not precluded from adding to the theory additional Majorana mass terms of the form
\begin{equation}
-\:\mathcal{L}\ \supset\ \frac{1}{2}\,\bar{\nu}_{R,\alpha}^{\mathcal{C}}m_{R,\alpha\beta}\nu_{R,\beta}\:+\:{\rm H.c.}\;.
\end{equation}
By including such terms and block diagonalising the full mass matrix, we obtain the non-Hermitian generalisation of the see-saw formula
\begin{equation}
\bm{m}_L\ =\ -\:\bm{m}_-\bm{m}_R^{-1}\bm{m}_+^{\mathsf{T}}\;.
\end{equation}
For $N=2$, the mass eigenvalues are
\begin{equation}
M_{1(2)}\ =\ -\:\frac{v^2}{4}\Big[{\rm tr}\,\bm{h}_-\bm{m}_R^{-1}\bm{h}_+^{\mathsf{T}}\:-\:(+)\Big(2\,{\rm tr}\big({\bm h}_-\bm{m}_R^{-1}{\bm h}_+^{\mathsf{T}}\big)^2\:-\:\big({\rm tr}\,\bm{h}_-\bm{m}_R^{-1}\bm{h}_+^{\mathsf{T}}\big)^2\Big)^{1/2}\Big]\;,
\end{equation}
and we trivially obtain a massless spectrum for $\bm{h}=\pm\,\bm{\eta}$. Instead, if
\begin{equation}
\mathrm{det}\,\bm{h}_-\bm{m}_R^{-1}\bm{h}_+^{\mathsf{T}}\ =\ 0\;,
\end{equation}
we obtain the spectrum
\begin{equation}
M_1\ =\ 0\;,\qquad M_2\ =\ -\:\frac{v^2}{2}\,{\rm tr}\,\bm{h}_-\bm{m}_R^{-1}\bm{h}_+^{\mathsf{T}}\;,
\end{equation}
which is real so long as
\begin{equation}
\mathrm{Im}\,{\rm tr}\,\bm{h}_-\bm{m}_R^{-1}\bm{h}_+^{\mathsf{T}}\ =\ 0\;.
\end{equation}
We can now obtain an arbitrarily small but finite mass splitting $\Delta M^2$ by arranging for
\begin{equation}
\mathrm{Re}\,{\rm tr}\,\bm{h}\bm{m}_R^{-1}\bm{h}^{\mathsf{T}}\ \gtrsim\ \mathrm{Re}\,{\rm tr}\Big(\bm{\eta}\bm{m}_R^{-1}\bm{\eta}^{\mathsf{T}}\:+\:\bm{h}\bm{m}_R^{-1}\bm{\eta}^{\mathsf{T}}\:-\:\bm{\eta}\bm{m}_R^{-1}\bm{h}^{\mathsf{T}}\Big)\;,
\end{equation}
and this mass splitting may be suppressed independent of the scale of the Majorana masses.

The anti-Hermitian terms are $\mathcal{C}\mathcal{P}$ odd, and one might naively anticipate additional sources of $\mathcal{C}\mathcal{P}$ violation. With the inclusion of the lepton number violating Majorana mass terms, it would therefore be interesting to consider this model in the context of leptogenesis and the generation of the Baryon Asymmetry of the Universe. Furthermore, the ability to suppress the neutrino masses independent of the Majorana mass scale may allow the see-saw scale to be lowered, and a full analysis of the constraints on the non-Hermitian Yukawa couplings from neutrino oscillation data and the current limits on lepton-flavour and lepton-number violating observables, including neutrinoless double beta decay, is warranted (for a review, see Ref.~\cite{Deppisch:2015qwa}). It would also be of interest to consider analogous non-Hermitian extensions of left-right symmetric models~\cite{Mohapatra:1974hk,Mohapatra:1974gc,Senjanovic:1975rk}. Finally, we remark that the consequences of the anti-Hermitian Yukawa couplings for the behaviour of the Higgs sector in the presence of fermion condensates remains to be studied.

\section{Concluding remarks}

We have described a $U(1)$ gauge theory coupled to a fermion with an anti-Hermitian, parity-violating mass term. We have shown that this model actually encompasses a continuum of theories, ranging from a massless left-Weyl theory to a massless right-Weyl theory, and we recover a massive Dirac fermion in the Hermitian limit. Moreover, we have illustrated that the behaviour of this theory is stable under loop corrections.

By constructing an analogous Higgs-Yukawa model, we have described how the presence of Hermitian and non-Hermitian Yukawa couplings of order unity can give rise to small neutrino masses, whilst at the same time leading to a suppression of the right-chiral component of the neutrino current. In and of itself, this model would not alleviate the need for fine-tuning in the SM, since we must arrange for the Hermitian and anti-Hermitian Yukawa couplings to differ only by one part in $10^{12}$. However, the small mass splitting needed to fit the observed neutrino spectrum could reasonably arise, for instance, through some radiative breaking of a high-scale degeneracy of the Hermitian and non-Hermitian Yukawa couplings. This point may be addressed in further work.\footnote{Particular thanks go to Andreas Trautner for emphasising the need to explain this relative fine-tuning of the Hermitian and non-Hermitian Yukawa couplings, as well as to Nikolaos Mavromatos for his constructive and encouraging remarks on this point.} With this goal in mind, it would be interesting to consider ways of generating the anti-Hermitian mass term, for example, through non-perturbative effects (see, e.g., Refs.~\cite{Alexandre:2013tya,Alexandre:2014zta,Alexandre:2012dm}).

\ack

The work of PM is supported by STFC grant ST/L000393/1. PM would like to thank the organisers of DISCRETE2016 for their hospitality and the opportunity to present this work, and the participants for their questions, comments and suggestions.

\end{document}